\documentclass[english,aps,manuscript]{revtex4}
\usepackage[T1]{fontenc}
\usepackage[latin1]{inputenc}
\usepackage{float}
\usepackage{graphicx}

\makeatletter

\providecommand{\tabularnewline}{\\}

\usepackage{babel}
\makeatother
\begin{document}

\title{Deviation from Bimaximality due to Planck Scale Effects }

\author{Bipin Singh Koranga and S. Uma Sankar}

\affiliation{Department of Physics, Indian Institute of Technology Bombay, Mumbai
400076, India.}

\author{Mohan Narayan}

\affiliation{Department of Physics, Mumbai University Institute of Chemical
Technology, Mumbai 400019, India.}

\begin{abstract}
We consider the effect of Planck scale operators on
neutrino mixing. We assume that GUT scale operators give
rise to degenerate neutrino masses with bimaximal mixing.
Quantum gravity (Planck scale) effects lead to
an effective $SU(2)_{L}\times U(1)$ invariant dimension-5 Lagrangian
involving neutrino and Higgs fields. 
This gives rise to additional terms in the neutrino mass matrix
on electroweak symmetry breaking.
These additional terms can be considered as a perturbation to the
GUT scale bi-maximal neutrino mass matrix. We assume that the gravitational
interaction is flavour blind and compute the deviations of the three
neutrino mixing angles due to the Planck scale effects. We find that the
changes in $\theta_{13}\,\,$and $\theta_{23}$ are very small but
the change in solar mixing angle $\theta_{12}$~can be as large as
$3.5^{o}$.
\end{abstract}
\maketitle

\section{Introduction}

One of the challenges in neutrino physics is to explain the pattern
of neutrino masses and mixings which are deduced from the current
neutrino data. Presently accepted values of the mixing angles are
$\theta_{12}=34^{o}\pm7^{o},$ $\theta_{13}\leq12^{o}$ and 
$\theta_{23}=45^{o}\pm9^{o}$ \cite{globalfits}.
It is reasonable to suppose that the symmetries of the Lagrangian
predict the elements of MNS matrix to be simple fractions. In a well-known
scenario called bi-maximal mixing \cite{bimaximal}, the mixing angles
are predicted to be $\theta_{12}=45^{o},$ $\theta_{13}=0^{o}$ and
$\theta_{23}=45^{o}.$ Additional effects can modify the above predictions
and lead to values of mixing angles close to experimentally determined
values. For example, it is well-known that renormalization group evolution
from GUT scale to electroweak scale can substantially change the value
of $\theta_{12}$ \cite{rgrunning}. The neutrino mass matrix is 
assumed to be generated by the see saw mechanism \cite{seesaw}. Here we will
assume that the dominant part of neutrino mass matrix arises due to
GUT scale operators and they lead to bi-maximal mixing. This matrix
can receive corrections due to physics from higher scale. 

The gravitational
interaction of neutrinos with the Standard Model Higgs field can be
expressed as an effective $SU(2)_{L}\times U(1)$ invariant dimension-5
operator \cite{weinberg},
\begin{equation}
L_{grav}=\frac{\lambda_{\alpha\beta}}{M_{pl}}(\psi_{A\alpha}\epsilon_{AC}
\psi_{C})C_{ab}^{-1}(\psi_{Bb\beta}\epsilon_{BD}\psi_{D})+h.c.
\end{equation}
Here and everywhere below we use Greek indices $\alpha,\beta$..for
the flavour states and Latin indices i, j, k for the mass states.
In the above equation $\psi_{\alpha}=(\nu_{\alpha},l_{\alpha})$ is
the lepton doublet~, $\phi=(\phi^{+},\phi^{0})$ is the Higgs doublet
and $M_{pl}=1.2\times10^{19}GeV$ is the Planck mass. $\lambda\,\,$is
a $3\times3$ matrix in flavour space with each element O(1). In eq(1),
all indices are explicitly shown. The Lorentz indices a, b = 1, 2,
3, 4 are contracted with the charge conjugation matrix C and the SU(2)L
isospin indices A, B, C, D = 1, 2 are contracted with $\epsilon,$
the Levi-Civita symbol in two dimensions. After spontaneous electroweak
symmetry breaking the Lagrangian in eq(1) generates additional terms
to the neutrino mass matrix
\begin{equation}
L_{mass}=\frac{v^{2}}{M_{pl}}\lambda_{\alpha\beta}\nu_{\alpha}C^{-1}\nu_{\beta},\end{equation}
where $v$=174 GeV is the VEV of electroweak symmetry breaking. 

We assume that gravitational interaction is {}``flavour blind'',
that is $\lambda_{\alpha\beta}\,\;$ is independent of 
$\alpha,\,\,\beta$~indices. This is a reasonable assumption because
gauge invariance requires gravity to couple to the spin of a particle
and these couplings are independent of the global $U(1)$ charges
of the particle. Thus the Planck scale contribution to the neutrino
mass matrix is
\begin{equation}
\mu\,\,\,\lambda=\mu\left(\begin{array}{ccc}
1 & 1 & 1\\
1 & 1 & 1\\
1 & 1 & 1\end{array}\right),
\end{equation}
where the scale $\mu$~is
\begin{equation}
\mu=\frac{v^{2}}{M_{pl}}=2.5\times10^{-6}eV.
\end{equation}
In our calculation, we take eq(3) as a perturbation to the main part
of the neutrino mass matrix that is generated by GUT dynamics. We
compute the changes in neutrino mass eigenvalues and mixing angles
induced by this perturbation.

\section{Corrections to mixing angles and neutrino mass squared differences}

We assume that GUT scale operators give rise to the light neutrino
mass matrix, which in the mass eigenbasis, takes the form $M=diag($
$M_{1},$$M_{2}$, $M_{3})$, where $M_{i}$ are real and non negative.
We take these to be the unperturbed ($0^{th}-order)$~masses. Let
$U$ be the neutrino mixing matrix at $0^{th}-order$. Then the corresponding
$0^{th}-order$ mass matrix $\mathbf{M}$ in flavour space is given
by
\begin{equation}
\mathbf{M}=U^{*}MU^{\dagger}.
\end{equation}

The $0^{th}-order$ MNS matrix $U$ has the form 
$\delta$\begin{equation}
U=\left(\begin{array}{ccc}
U_{e1} & U_{e2} & U_{e3}\\
U_{\mu1} & U_{\mu2} & U_{\mu3}\\
U_{\tau1} & U_{\tau2} & U_{\tau3}\end{array}\right),
\end{equation}
where the nine elements are, in general, functions of three mixing angles
and six phases. In terms of the above elements,
the mixing angles are defined by 
\begin{eqnarray}
|\frac{U_{e2}}{U{}_{e1}}| & = & tan\theta_{12},  \\
|\frac{U_{\mu3}}{U_{\tau3}}| & = & tan\theta_{23}, \\
|U_{e3}| & = & sin\theta_{13}.
\end{eqnarray}
In terms of the above mixing angles, the mixing matrix is written
as 
\begin{equation}
U=diag(e^{if1},\,\, e^{if2},\,\, e^{if3})R(\theta_{23})\Delta R(\theta_{13})\Delta^{*}R(\theta_{12})diag(e^{ia1},e^{ia2},1).\end{equation}
The matrix $\Delta=diag(e^{\frac{i\delta}{2}},1,e^{\frac{-i\delta}{2}})$
contains the Dirac phase $\delta.$ This phase leads to CP violation
in neutrino oscillations. $a1$ and $a2$~are the so called Majorana
phases, which affect the neutrinoless double beta decay. $f1,\,\, f2$
and $f3$~are usually absorbed as a part of the definition of the
charged lepton field. It is possible to rotate these phases away,
if the mass matrix in eq(5) is the complete mass matrix. However,
since we are going to add another contribution to this mass matrix,
these phases of the zeroth order mass matrix can have an impact on
the complete mass matrix and thus must be retained. By the same token,
the Majorana phases which are usually redundant for oscillations have
a dynamical role to play now. Planck scale effects will add other
contributions to the mass matrix. Including the Planck scale mass
terms, the mass matrix in flavour space is modified as
\begin{equation}
\mathbf{M}\rightarrow\mathbf{M^{'}=M}+\mu\lambda,\end{equation}
with $\lambda$ being a matrix whose elements are all 1 as discussed
in eq(3). Since $\mu$ is small, we treat the second term (arising
from the Planck scale effects) in the above equation as a perturbation
to the first term (the GUT scale mass terms). This perturbation formalism
was first developed in ref. \cite{vissani}. Here we briefly recall the main
features for completeness. The matrix relevant for oscillation physics
is the following hermitian matrix
\begin{equation}
\mathbf{M^{'^{\dagger}}M^{'}=}(\mathbf{M}+\mu\lambda)^{\dagger}(\mathbf{M}+\mu\lambda).\end{equation}
To the first order in the small parameter $\mu$, the above matrix
is
\begin{equation}
\mathbf{M}^{\dagger}\mathbf{M}+\mu\lambda^{\dagger}\mathbf{M}+\mathbf{M}^{\dagger}\mu\lambda.\end{equation}
This hermitian matrix is diagonalized by a new unitary matrix $U^{'}.$
The corresponding diagonal matrix $M^{'^{2}},$ correct to first order
in $\mu$, is related to the above matrix by $U^{'}M^{'^{2}}U^{'^{\dagger}}.$
Rewriting $\mathbf{M}$ in the above expression in terms of the diagonal
matrix $M$ we get 
\begin{equation}
U^{'}M^{'^{2}}U^{'^{\dagger}}=U(M^{2}+m^{\dagger}M+Mm)U^{\dagger}\end{equation}
where \begin{equation}
m=\mu U^{t}\lambda U.\end{equation}
Here $M$ and $M^{'}$ are the diagonal matrices with neutrino masses
correct to $0^{th}\,\,$and $1^{th}\,\,$order in $\mu$. It is clear
from eq(14) that the new mixing matrix can be written as:
\begin{equation}
U^{'}=U(1+i\delta\Theta),\end{equation}
where $\delta\Theta$~is a hermitian matrix that is first order in
$\mu.$ Oscillation physics is unchanged under the transformation
$U\rightarrow UP$, where $P$ is a diagonal phase matrix. We can use
this invariance to set the diagonal elements of the matrix $\delta\Theta$
to be zero.

From eq(14) we obtain
\begin{equation}
M^{2}+m^{\dagger}M+Mm=M^{'^{'2}}+[i\delta\Theta,M^{'^{2}}].\end{equation}
Therefore to first order in $\mu$,~the mass squared difference $\Delta M_{ij}^{2}=M_{i}^{2}-M_{j}^{2}$~get
modified as
\begin{equation}
\Delta M_{ij}^{'^{2}}=\Delta M_{ij}^{2}+2(M_{i}Re[m_{ii}]-M_{j}Re[m_{jj}]).\end{equation}
The change in the elements of the mixing matrix, which we parametrized
by $\delta\Theta$, is given by
\begin{equation}
\delta\Theta_{ij}=\frac{iRe(m_{ij})(M_{i}+M_{j})}{\Delta M_{ij}^{'^{2}}}-\frac{Im(m_{ij})(M_{i}-M_{j})}{\Delta M_{ij}^{'^{2}}}.\end{equation}
The above equation determines only the off diagonal elements of matrix
$\delta\Theta_{ij}$. As mentioned above, the diagonal elements can
be set to zero without loss of generality. The expressions for $\Delta M_{ij}^{'2}$
in eq(18) and for $\delta\Theta_{ij}$ in eq(19) were first obtained
in ref \cite{vissani}. From the above equation, we see that~$\delta\Theta_{ij}$
are proportional to the neutrino masses. Thus they are larger for
the case of degenerate neutrinos. Here onwards we assume degenerate
neutrino masses. Then expression for $\delta\Theta_{ij}$ simplifies
to 
\begin{equation}
\delta\Theta_{ij}=\frac{iRe(m_{ij})}{M_{i}-M_{j}}-\frac{Im(m_{ij})}{M_{i}+M_{j}}.\end{equation}
In obtaining the above equation, we made the approximation $\Delta M_{ij}^{'^{2}}=\Delta M_{ij}^{2}.$
This is valid because the $m_{ij}$~term in the numerator is already
proportional to $\mu$ and we are working to $1^{st}$ order in $\mu$.
For degenerate neutrinos, $M_{i}+M_{j}\gg M_{i}-M_{j}.$ Thus the
expression for $\delta\Theta_{ij}$ further simplifies to
\begin{equation}
\delta\Theta_{ij}=\frac{iRe(m_{ij})}{M_{i}-M_{j}}.\end{equation}

To obtain the largest effect possible, we assume the largest allowed
value of 2 eV for degenerate neutrino mass which comes from tritium
beta decay \cite{tritium}. We also assume normal neutrino mass hierarchy.
Thus we have $M_{1}=$2 eV, $M_{2}=\sqrt{M_{1}^{2}+\Delta_{21}}$
and $M_{3}=\sqrt{M_{1}^{2}+\Delta_{31}}$. If all elements of $\lambda$~consist
of 1 then $m_{ij}=z_{i}z_{j},$~where three complex numbers $z_{i}$
are defined by
\begin{equation}
z_{1}=U_{e1}+U_{\mu1}+U_{\tau1},\end{equation}
\begin{equation}
z_{2}=U_{e2}+U_{\mu2}+U_{\tau2},\end{equation}
\begin{equation}
z_{3}=U_{e3}+U_{\mu3}+U_{\tau3}.\end{equation}
As in the case of $0^{th}$order mixing angles, we can define $1^{st}$
order mixing angles in terms of $1^{st}$ order mixing matrix elements
in a manner similar to eq(19)
\begin{equation}
tan\theta_{12}^{'}=|\frac{U_{e2}^{'}}{U_{e1}^{'}}|\end{equation}
\begin{equation}
tan\theta_{23}^{'}=|\frac{U_{\mu3}^{'}}{U_{\tau3}^{'}}|\end{equation}
\begin{equation}
sin\theta_{13}^{'}=|U_{e3}^{'}|.\end{equation}
From eq(16), we get
\begin{equation}
\delta U_{\alpha j}=U_{\alpha j}^{'}-U_{\alpha j}=
i \sum_{i=1}^{3} U_{\alpha i}\delta\Theta_{ij}
\end{equation}
Substituting the expressions for $\delta\Theta_{ij}$ from eq(21) 
(and $\delta \Theta_{ii} =0$) in the above equation, we obtain
\begin{equation}
\delta U_{e1}=\mu\left(U_{e2}\frac{Re(z_{1}z_{2})}{M_{2}-M_{1}}+U_{e3}\frac{Re(z_{1}z_{3})}{M_{3}-M_{1}}\right),\end{equation}
\begin{equation}
\delta U_{e2}=\mu\left(U_{e1}\frac{Re(z_{1}z_{2})}{M_{2}-M_{1}}+U_{e3}\frac{Re(z_{2}z_{3})}{M_{3}-M_{2}}\right),\end{equation}
\begin{equation}
\delta U_{e3}=\mu\left(U_{e1}\frac{Re(z_{1}z_{3})}{M_{3}-M_{1}}+U_{e2}\frac{Re(z_{2}z_{3})}{M_{3}-M_{2}}\right),\end{equation}
\begin{equation}
\delta U_{\mu3}=\mu\left(U_{\mu1}\frac{Re(z_{1}z_{3})}{M_{3}-M_{1}}+U_{\mu2}\frac{Re(z_{2}z_{3})}{M_{3}-M_{2}}\right),\end{equation}
\begin{equation}
\delta U_{\tau3}=\mu\left(U_{\tau1}\frac{Re(z_{1}z_{3})}{M_{3}-M_{1}}+U_{\tau2}\frac{Re(z_{2}z_{3})}{M_{3}-M_{2}}\right).\end{equation}

For degenerate neutrinos, $M_{3}-M_{1}\cong M_{3}-M_{2}\gg M_{2}-M_{1}$
because $\Delta_{31}\cong\Delta_{32}\gg\Delta_{21}.$ Thus, from the
above set of equations, we see that $\delta U_{e1}$ and $\delta U_{e2}$
are much larger than $\delta U_{e3},\,\,\,$$\delta U_{\mu3}$ and
$\delta U_{\tau3}$. Hence we can expect much larger change in $\theta_{12}$
compared to $\theta_{13}$ and $\theta_{23}$.

\section{numerical Results}

As mentioned in the introduction, we expect the mixing angles coming
from GUT scale operators to be determined by some symmetries. 
We assume these symmetries give riso to a bi-maximal mixing pattern, 
$\theta_{12}=\theta_{23}=\pi/4$
and $\theta_{13}=0$. Using eq(25) to eq(27), we compute the modified
mixing angles for the degenerate neutrino mass of 2 eV.~ We have
taken $\Delta_{31}=0.002eV^{2}$ \cite{superk} and $\Delta_{21}=0.00008eV^{2}$
\cite{kamland}. For simplicity, we have set the charged lepton phases
$f1=f2=f3=0$. Since we have set $\theta_{13}=0$, the Dirac phase
$\delta$ drops out of the $0^{th}$ order mixing matrix. From eq(22)
to eq(24) we see that the complex numbers $z_{1},$~$z_{2}$~~and
$z_{3}$ are independent of $\delta$ and hence all deviations in $U_{\alpha j}$
are independent of $\delta$. The complex nature of $z_{1},$~$z_{2}\,\,\,$and
$z_{3}$ comes from the Majorana phases $a1$ and $a2$, which we
take to be non-zero. We compute the modified mixing angles as function
of $a1$ and $a2$. In table 1, we list the modified neutrino mixing
angles for some sample values of $a1$ and $a2$. As shown in the
table, the deviation in $\theta_{13}$ and $\theta_{23}$ is negligibly
small whereas the deviation in $\theta_{12}$ is significant. In fig(1),
we show contours of constant deviation, $\delta \theta_{12} = 
\theta_{12}^{'}-\theta_{12})$, ~vs $a1$~and $a2$.

\begin{figure}[H]
~~~~\includegraphics[%
  scale=0.5,
  angle=-90]{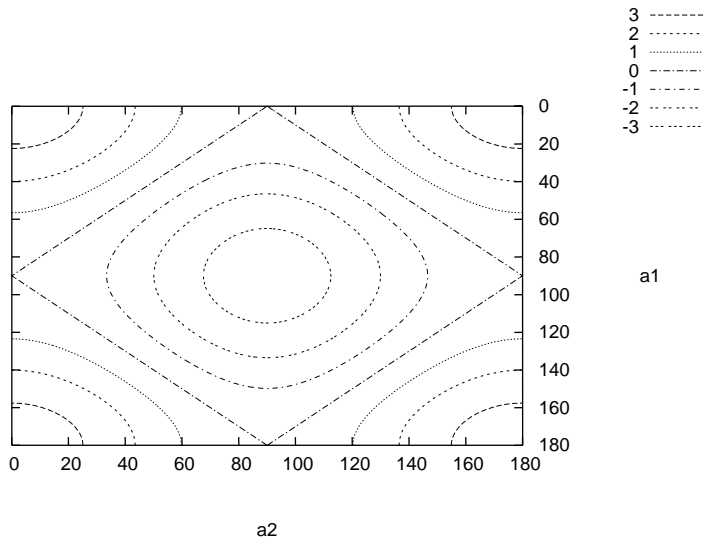}

\caption{Deviation of $\theta_{12}$~as a function of the Majorana phases
in the case of bi-maximal mixing.}
\end{figure}

\begin{figure}[H]
~~~~\includegraphics[%
  scale=0.5,
  angle=-90]{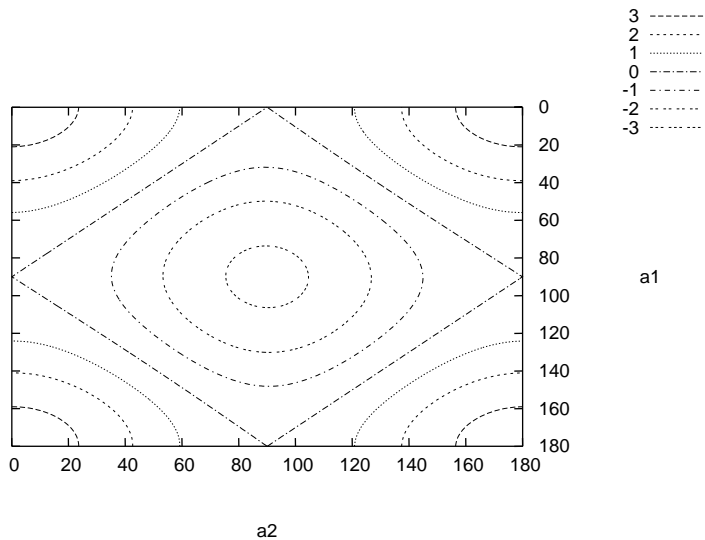}

\caption{Deviation of $\theta_{12}$~as a function of the Majorana phases
in the case of tri-bi-maximal mixing.}
\end{figure}

From fig(1), we see that Planck scale effects reduce $\theta_{12}$
from the bi-maximal value of $45^{o}$ to about $\theta_{12}^{'}=41.5^{o}.$
This is the present 3$\sigma$ upper limit of the solar mixing angle.
Thus we see that Planck scale effects can bring down bi-maximal $\theta_{12}$
to within experimentally acceptable range. In our discussion above,
we have not included the renormalization group running. The matrix
in eq(3) arises at Planck scale. Its renormalization group running from
$M_{pl}$ to $M_{GUT}$ is negligible. Thus, at $M_{GUT}$, the total
interaction matrix, that eventually gives rise to neutrino mass matrix
on electroweak symmetry breaking, is the sum of GUT scale matrix and
the Planck scale matrix. Renormalization group running modfies these couplings
from their GUT values to low energy values \cite{rgrunning}.
This can lead to lower values of $\theta_{12}^{'}$, much closer to
the experimental best fit value. 

Recently various authors considered tri-bi-maximal mixing scenario
which predicts $sin^{2}\theta_{12}=1/3$ or $\theta_{12}=35^{o},$
in order to obtain a value close to the best fit value \cite{tribimax}. 
We computed the modified mixing angles for this scenario
also, that is for the input values $\theta_{12}=35^{o},$$\theta_{13}=0^{o}$and
$\theta_{23}=45^{o}.$ Neutrino masses are kept to be the same as
in the previous case. The modified mixing angles are shown in table
(2). In fig (2) the deviation of $\theta_{12}$is plotted as a function
of $a1$~and $a2.$ We see that deviation in $\theta_{12}$ is about
$\pm3^{o}$. The range of $\theta_{12}$ given by this deviation spans
$1\sigma$ range of solar neutrino mixing angle. 

\begin{table}[H]
~~~~~~\begin{tabular}{|c|c|c|c|c|}
\hline 
~~~~ $a1$~~~~&
~~~~$a2$~~~~&
~~~~$\theta_{12}^{'}$~~~~&
~~~~$\theta_{23}^{'}$~~~~ &
~~~~$\theta_{13}^{'}$~~~~\tabularnewline
\hline
\hline 
0&
0&
48.57&
45.00&
0.28\tabularnewline
\hline 
0&
45&
46.87&
44.93&
0.22\tabularnewline
\hline 
0&
90&
44.99&
44.86&
0.13\tabularnewline
\hline 
0&
135&
46.94&
44.93&
0.22\tabularnewline
\hline 
0&
180&
48.57&
45.00&
0.28\tabularnewline
\hline 
45&
0&
46.68&
45.07&
0.22\tabularnewline
\hline 
45&
45&
44.96&
45.00&
0.20\tabularnewline
\hline 
45&
90&
43.09&
44.93&
0.09\tabularnewline
\hline 
45&
135&
45.03&
45.00&
0.14\tabularnewline
\hline 
45&
180&
46.68&
45.07&
0.22\tabularnewline
\hline 
90&
0&
45.00&
45.14&
014\tabularnewline
\hline 
90&
45&
43.28&
45.07&
0.10\tabularnewline
\hline 
90&
90&
41.42&
45.00&
0.00005\tabularnewline
\hline 
90&
135&
43.00&
45.07&
0.10\tabularnewline
\hline 
90&
180&
45.00&
45.14&
0.14\tabularnewline
\hline 
135&
0&
46.68&
45.07&
0.22\tabularnewline
\hline 
135&
45&
44.96&
45.00&
0.14\tabularnewline
\hline 
135&
90&
43.09&
44.93&
0.09\tabularnewline
\hline 
135&
135&
45.03&
45.00&
0.20\tabularnewline
\hline 
135&
180&
46.68&
45.07&
0.22\tabularnewline
\hline 
180&
0&
48.57&
45.00&
0.22\tabularnewline
\hline 
180&
45&
46.87&
44.93&
0.22\tabularnewline
\hline 
180&
90&
44.99&
44.86&
0.13\tabularnewline
\hline 
180&
135&
46.94&
44.93&
0.22\tabularnewline
\hline 
180&
180&
48.57&
45.00&
0.28\tabularnewline
\hline
\end{tabular}

\caption{The modified mixing angles for various values of phases. Input values
are $\Delta_{31}=0.002eV^{2}$,~$\Delta_{21}=0.00008eV^{2}$ , $\theta_{12}=\theta_{23}=45^{o}$,~$\theta_{13}=0^{o}$.}
\end{table}

\begin{table}[H]
~~~~~~\begin{tabular}{|c|c|c|c|c|}
\hline 
 ~~~~$a1$~~~~&
~~~~$a2$~~~~&
~~~~$\theta_{12}^{'}$~~~~&
~~~~$\theta_{23}^{'}$ ~~~~&
~~~~$\theta_{13}^{'}$~~~~\tabularnewline
\hline
\hline 
0&
0&
38.51&
45.00&
0.28\tabularnewline
\hline 
0&
45&
36.82&
44.93&
0.23\tabularnewline
\hline 
0&
90&
34.99&
44.86&
0.18\tabularnewline
\hline 
0&
135&
 36.88&
 44.93&
 0.23\tabularnewline
\hline 
0&
180&
38.51&
45.00&
0.28\tabularnewline
\hline 
45&
0&
36.63&
45.06&
0.21\tabularnewline
\hline 
45&
45&
34.97&
45.00&
0.19\tabularnewline
\hline 
45&
90&
 33.26&
44.93&
0.13\tabularnewline
\hline 
45&
135&
 35.09&
45.00&
0.15\tabularnewline
\hline 
45&
180&
 36.63&
45.06&
0.21\tabularnewline
\hline 
90&
0&
 35.00&
45.13&
0.09\tabularnewline
\hline 
90&
45&
 33.43&
45.06&
0.06\tabularnewline
\hline 
90&
90&
31.77&
45.00&
0.00005\tabularnewline
\hline 
90&
135&
33.49&
45.06&
0.05\tabularnewline
\hline 
90&
180&
35.00&
45.13&
0.09\tabularnewline
\hline 
135&
0&
36.63&
 45.06&
0.21\tabularnewline
\hline 
135&
45&
 35.04&
45.00&
 0.14\tabularnewline
\hline 
135&
90&
33.26&
44.93&
0.13\tabularnewline
\hline 
135&
135&
35.02&
 45.00&
0.19\tabularnewline
\hline 
135&
180&
 36.63&
 45.06&
 0.21\tabularnewline
\hline 
180&
0&
38.51&
45.00&
0.28\tabularnewline
\hline 
180&
45&
36.82&
44.94&
0.24\tabularnewline
\hline 
180&
90&
 34.99&
44.86&
0.18\tabularnewline
\hline 
180&
135&
 36.88&
44.94&
 0.24\tabularnewline
\hline 
180&
180&
 38.51&
45.00&
0.28\tabularnewline
\hline
\end{tabular}

\caption{The modified mixing angles for various values of phases. Input values
are |$\Delta_{31}|=0.002eV^{2}$,~$\Delta_{21}=0.00008eV^{2}$ ,
$\theta_{12}=35^{o},\,\,\theta_{23}=45^{o}$,~$\theta_{13}=0^{o}$.}
\end{table}

\section{conclusions}

It is expected that symmetries at GUT scale will determine the neutrino
mixings. Bi-maximal mixing is one of the attractive theoretical ideas
proposed in this context. However, the solar neutrino data show that
the mixing angle $\theta_{12}$ is substantially less than $\pi/4$.
In this paper, we studied how neutrino mass terms arising from Planck
scale effects modify the bi-maximal mixing scenario. We consider
these additional mass terms to be perturbation to the main neutrino
mass matrix coming from GUT scale and computed the first order corrections
to neutrino masses and mixings. The changes in all three mixing angles
are proportional to the neutrino mass eigenvalues. To maximise the
change we assumed degenerate neutrino masses $\simeq2.0$~eV. For
degenerate neutrino masses, the changes in 
$\theta_{13}$~and $\theta_{23}\,\,$are
inversely proportional to $\Delta_{31}$ and $\Delta_{32}$ respectively,
whereas the change in $\theta_{12}$ is inversely proportional to
$\Delta_{21}$. Since $\Delta_{31}\cong\Delta_{32}\gg\Delta_{21},$
the change in $\theta_{12}$ is much larger than the changes in 
$\theta_{13}\,\,$and
$\theta_{23}$. The actual magnitude of change also depends on the
Majorana phases $a1$ and $a2$ but is independent of Dirac phase
$\delta$, because we assumed that $\theta_{13}$ at $0^{th}$ order
is zero. In ref. \cite{vissani}, it was shown that the change in $\theta_{13}$,
due to this perturbation, is small. Here we show that the change in
$\theta_{23}$ also is small (less than $0.3^{o})$ but the change
in $\theta_{12}\,\,$can be substantial (about $\pm3^{o})$. 
If $\theta_{12}\,\,$at
$0^{th}$ order is $45^{o},$ the change induced by the Planck scale
terms can bring it down to $41.5^{o},$ which is the $3\sigma$ upper
limit of the solar mixing angle.


\begin{thebibliography}{10}
\bibitem{globalfits}
G.L. Fogli, E. Lisi, A. Marrone and A. Palazzo, Prog. Part. Nucl. Phys.
{\bf 57}, 742-795, (2006);
A. Strumia and F. Vissani, Nucl. Phys. {\bf B 726}, 294 (2005);
S. Goswami, A. Bandyopadhyaya and S. Chobey, 
Nucl. Phys. Proc. Suppl. {\bf 143}, 121 (2005);
P. C. de Holanda and A. Yu. Smirnov, Astropart. Phys. {\bf 21}, 287 
(2004).
\bibitem{bimaximal}
Nan Li and Bo-Qiang Ma, Phys.Rev.\textbf{D71}, 017302, 2005;
C.A. de S. Pires, J.Phys.\textbf{G30}, B29, (2004);
P.H. Frampton, S.T. Petcov and W. Rodejohann, 
Nucl.Phys.\textbf{B687},31-54 (2004).
\bibitem{rgrunning}
S. Pokorski, Int. J. Mod. Phys. {\bf A17}, 575 (2002);
M. Lindner, Nucl. Phys. {\bf B 674}, 401 (2003) and the references therein.
\bibitem{seesaw}
P. Minkowski, Phys. Lett. {\bf B67}, 421 (1977);
M. Gell-Mann, P. Ramond and R. Slanksky, in "Supergravity" Stonybrook, 1979,
eds. D. Freedman and P. van Nieuwenhiuzen; 
T. Yanagida, in "Proceedings of the workshop on unified theory and 
baryon number in the universe" KEK, March 1979, eds. O. Sawada and 
A. Sugamoto;  
R. N Mohapatra and G. Senjanovic, Phys. Rev. Lett. \textbf{44}, 912 (1980);
R. N Mohapatra and G. Senjanovic, Phys. Rev. D \textbf{23}, 165 (1981).
\bibitem{weinberg}
S. Weinberg, Phys. Rev. Lett. \textbf{43}, 1566 (1979).
\bibitem{vissani}
F. Vissani, M. Narayan and V. Berezinsky, Phys. Lett. \textbf{B 571},
209, 2003.
\bibitem{tritium}
C. Kraus $et\,\,\, al,.$ Eur. Phys. J. {\bf C 33}, S805 (2004).
V. M. Lobashev $et\,\,\, al,.$ Nucl. Phys. Proc. Suppl. {\bf 91}, 280 
(2001).
\bibitem{superk}
Super-Kamiokande Collaboration, J. Hosaka {\it et al}, 
Phys. Rev. D {\bf 74}, 032002 (2006).
\bibitem{kamland}
KamLAND Collaboration, T. Araki {\it et al}, Phys. Rev. Lett. {\bf 94},
081801 (2005).
\bibitem{tribimax}
P. F. Harrison, D. H. Perkins and W. G. Scott, Phys. Lett. \textbf{B}
530, (2002) 167;
P. F. Harrison and W. G. Scott, Phys. Lett. \textbf{B} 557, (2003)
76;
I. de M. Varzielas, S. F. King and G. G. Ross, arXiv: hep-ph/0607045.
\end{thebibliography}
\end{document}